\documentclass[aps,prb,reprint,superscriptaddress,amsmath,amssymb,floatfix]{revtex4-2}

\usepackage{graphicx}
\usepackage{dcolumn}
\usepackage{bm}
\usepackage{hyperref}
\usepackage[version=3]{mhchem}
\usepackage{physics}
\usepackage{braket}
\usepackage{siunitx}
\usepackage[normalem]{ulem}
\usepackage{chemformula}

\hypersetup{
    colorlinks=true,
    linkcolor=blue,
    filecolor=magenta,      
    urlcolor=cyan,
    citecolor=blue,
}

\begin{document}

\title{Emergent Hierarchy in Localized States of Organic Quantum Chains}

\author{L. L. Lage}
\affiliation{Instituto de Física, Universidade Federal Fluminense, Av. Litorânea s/n, Niterói, 24210-340, RJ, Brazil}

\author{A. B. Félix}
\affiliation{Instituto de Física, Universidade Federal Fluminense, Av. Litorânea s/n, Niterói, 24210-340, RJ, Brazil}

\author{D. S. Gomes}
\affiliation{Gleb Wataghin Institute of Physics, Department of Applied Physics, State University of Campinas, Rua Sérgio Buarque de Holanda, 777, Campinas, 13083-859, SP, Brazil}

\author{M. L. Pereira Jr.}
\affiliation{Department of Electrical Engineering, College of Technology, University of Brasília, Brasília, 70910-900, DF, Brazil}

\author{A. Latgé}
\email{andrea.latge@gmail.com}
\affiliation{Instituto de Física, Universidade Federal Fluminense, Av. Litorânea s/n, Niterói, 24210-340, RJ, Brazil}

\date{\today}

\begin{abstract}
Organic Quantum Chains (OQCs) represent a newly synthesized class of carbon-based nanostructures whose quasi-one-dimensional nature gives rise to unconventional electronic and transport phenomena. Here we investigate the electronic and transport properties of recently synthesized OQCs [Nature Communications, 12, 5895 (2021)]. Structural stability was first assessed through molecular dynamics relaxation combined with density functional theory (DFT). The optimized coordinates are then used in a tight-binding model with exponentially decaying hopping parameterization, which reproduces the DFT results with high accuracy. Our calculations reveal a robust and nearly constant energy gap across several OQC configurations, in agreement with experimental data. We also identify emergent hierarchical states, characterized by distinct localization behaviors within sets of localized bands. Finally, we analyze different transport responses in scenarios involving the one-dimensional OQC coupled to carbon corrals, as observed in the experimental data, highlighting their potential as promising systems for application in carbon nanodevices.
\end{abstract}

\maketitle

\section{Introduction}

Organic Quantum Chains (OQC) are a class of carbon-based systems that exhibit peculiar electronic and structural properties due to their quasi-one-dimensional architecture \cite{Nature2021, Paschke2025, Shu2018,Zhao2024}. These systems are composed of conjugated organic molecules arranged in a chain-like fashion, enabling efficient charge transport and tunable electronic behavior. The properties of OQCs make them promising candidates for applications in molecular electronics, optoelectronics, and quantum-spin computing \cite{Bhat2023, Fabian2025, Dubin2006, Zhao2024, Su2025, Zhao2025,doi:10.1021/jacs.5c03736}. Such low-dimensional systems have been synthesized employing self-assembly within chemical routes and direct manipulation with scanning microscope tips \cite{Zhong2021}. 
\begin{figure}[h!]
\centering
\includegraphics[width=8cm]{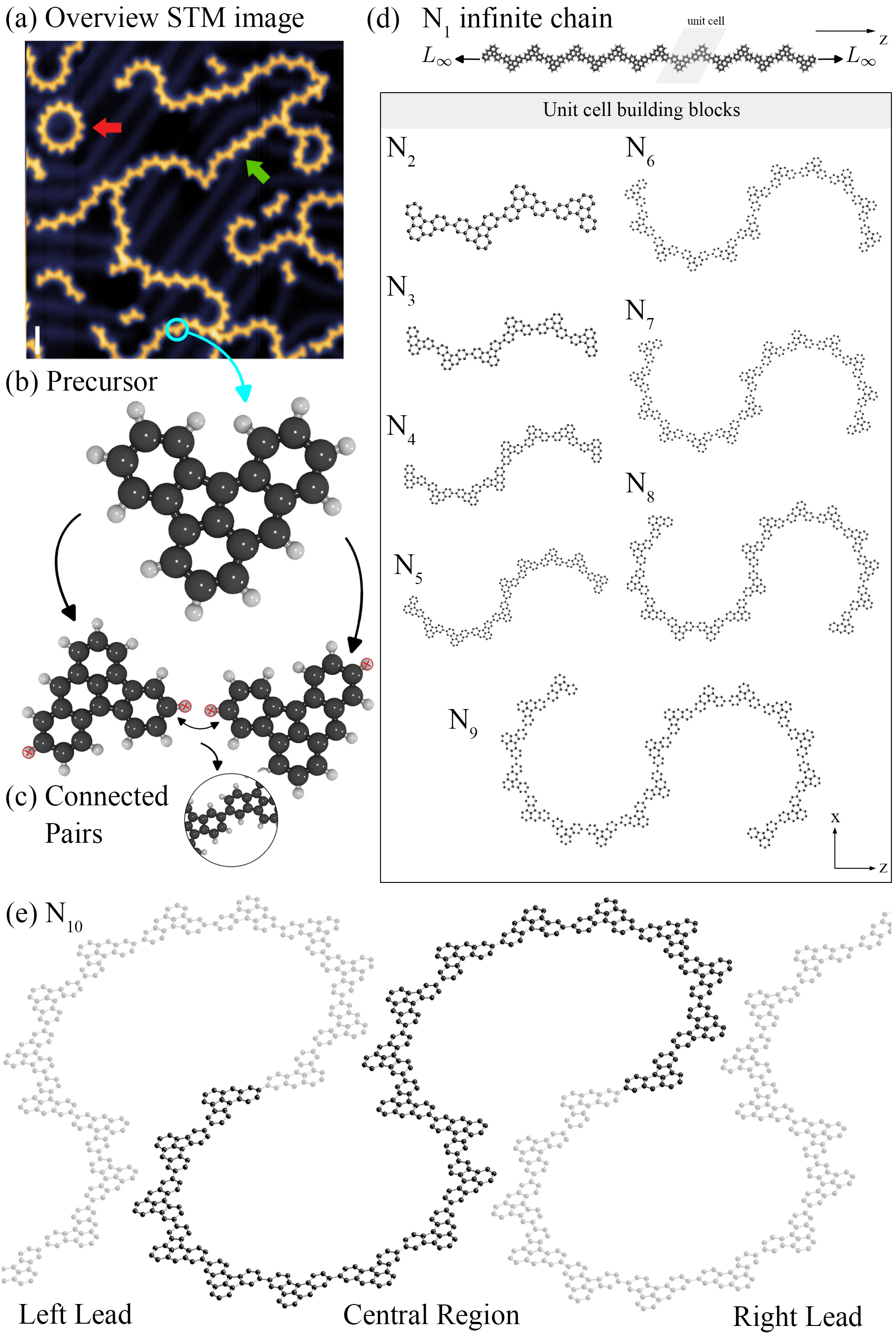}
\caption{(a) STM image highlighting: ring-like structures (red arrow) and chain segments (green arrow) [Adapted \cite{Nature2021}]. (b) Precursor used in the OQCs. (c) Connected pair of precursors. (d) Infinite chain composed of $N_1$-OQCs (unit cell marked in gray), and unit cell buildings blocks for $N_{2-9}$-OQCs (rectangular panel) illustrating the structural evolution and chain complexity along the periodic z-direction.(e) Scattering setup for $N_{10}$ with left and right leads.}
\label{fig:setup}
\end{figure}
 
Moreover, when constructed exclusively from carbon-based molecular building blocks (e.g., polycyclic aromatic hydrocarbons, graphene nanoribbons, or conjugated polymers), these systems exhibit emergent electronic localization, topological, and magnetic properties absent in bulk materials \cite{Kempkes2023,Chen2024,Schwab2015,PereiraJunior2020}. Very recently, nanoscaled chains were synthesized during a growth process of organic molecules on a gold substrate \cite{Nature2021}.

Stable macrocycle quantum corrals composed of twelve molecular units, marked with a red arrow in Fig.~\ref{fig:setup}\textcolor{black}{(a)}, are also generated, providing new designed quantum states. Such structures have a precursor unit composed of four hexagons and one pentagon, as depicted in Fig.~\ref{fig:setup}(b). In the sequence, Fig.~\ref{fig:setup}(c) shows the building pair process in the construction of an OQC, exemplified in Fig.~\ref{fig:setup}(d) as an N$_1$-OQC. In this work, we employ various theoretical frameworks for modeling OQCs, with a focus on their electronic structure, quantum transport properties, and potential applications. By combining DFT with Green's function formalism, we provide a detailed understanding of the fundamental properties of OQC and their implications for next-generation molecular devices.

\textcolor{black}{Low-dimensional organic systems are characterized by a unique interplay between electron delocalization, driven by $\pi$-orbitals, and spatial confinement arising from the atomic structure of the units. Unlike simple atomic chains, Organic Quantum Chains, such as those reported~\cite{Nature2021}, have large unit cells. The structural complexity gives rise to rich electronic spectra in which dispersive bands that support charge transport coexist with confined molecular states, manifested as highly localized, narrow-band modes. These states arise not from lattice topology, as in Kagomé and Lieb lattices \cite{Slot2017,Chen2024,HU201893,Kempkes2023,dicelattice,flatcompact}, or in higher-dimensional flat band systems \cite{Duan2024,Wang2024},
but from the specific potential confinement of the molecules raised by an increasing number of precursor units in the unit cell which brings dispersionless states. The electronic probability density in such narrow-band states remains localized within specific regions of the chain.} 


Typically, the coupling of finite molecules to continuum systems results in Fano antiresonance phenomena \cite{Hsu2016, Figueroa2024,FELIX2019}, characterized by a dip in transmission responses due to destructive interference. On the other hand, in the cases where molecular states are completely decoupled from the chain system, perfectly localized modes are formed inside the continuum state, named a bound quantum state in the continuum (BIC), without a change in the conductance value related to the decoupled system \cite{PhysRevA.106.013719,Martinez2025}. BICs are also discussed in wave problems such as acoustics, microwaves and nanophotonics \cite{MartiSabate2024,PhysRevLett.128.084301,PhysRevLett.107.183901}. In this work, we have investigated the emergence of localized states of different natures, creating a unique set of properties that can be directly observed in electronic responses. Understanding localized states is crucial in designing new devices that harness the robustness and quantum interference of these unique states.

\section{Theory \& Computational details}

\subsection{Systems Modeling} 
\textcolor{black}{To model the OQC formation process observed in the scanning tunneling microscope (STM) experiments~\cite{Nature2021}, we constructed the organic chains by covalently linking $C_{20}H_{12}$ precursors. The formation logic is detailed in Fig.~\ref{fig:setup} (b-c). Specifically, we removed the two outer hydrogen atoms located at the extremities of the precursor's longitudinal axis, highlighted with red markers in Fig.~\ref{fig:setup} (b). This dehydrogenation activates the terminal carbon sites, promoting the C-C coupling that results in the linear chain propagation observed experimentally.} The precursor molecule was then replicated along the $z$ axis, maintaining its orientation within the $xz$ plane. The replicated unit underwent sequential rotations about the $x$ and $z$ axes. The molecular units were positioned at an appropriate distance to enable the formation of covalent bonds between the terminal carbon atoms, and a vacuum region of \SI{100}{\AA} was added in the $x$ and $y$ directions. The lattice vector along the $z$ direction was adjusted to allow the formation of covalent bonds between the opposite ends of the replicated structure under periodic boundary conditions. The resulting dimer, when infinitely replicated in one direction, constitutes a one-dimensional system here referred to as N$_1$-OQC.

According to the experimental report, the precursor molecule tends to exhibit natural curvature, a feature exploited in the context of electronic confinement in corral-like structures \cite{Nature2021}. Based on this behavior, a molecular intercalation strategy was adopted to enable the formation of periodic one-dimensional systems with enhanced geometric stability. In this approach, N$_i$-OQC is defined as a system whose unit cell contains a pair of $i$ covalently connected molecules arranged in sequence. This configuration promotes mutual compensation of local curvature between adjacent molecular chains of the same length, contributing to the overall structural stability. In the present study, values of $i$ ranging from 1 to 10 were investigated. The number of carbon sites $(\mathcal{N})$ in the unit cell is given by $\mathcal{N}=40i$, where $i$ is the OQC index. The corresponding unit cells contained between 60 and 600 atoms (carbon and hydrogen), depending on the number of molecules.

The initial geometries were constructed in a linear configuration and did not account for the natural curvature of the OQCs. Given the high computational cost of structural optimization using DFT, an intermediate relaxation stage was implemented using classical molecular dynamics (CMD), carried out with the Large-scale Atomic/Molecular Massively Parallel Simulator (LAMMPS) \cite{plimpton1995fast}. The structural preparation protocol was divided into five stages to generate classically optimized geometries, which were then used as input for subsequent DFT calculations.

The first stage involved the simultaneous minimization of potential energy and internal forces, excluding thermal effects. The total energy was minimized using the conjugate gradient algorithm. The cell volume was also relaxed under zero pressure in the $z$ direction. To escape from local energy minima, a simulation in the NPT ensemble was performed for \SI{50}{ps} at a temperature of \SI{300}{K} and zero pressure along the longitudinal direction. This step induced realistic structural curvatures along the molecular chains. The system was then linearly cooled from \SI{300}{K} to \SI{1}{K} over \SI{25}{ps} in the same ensemble. An additional simulation of \SI{25}{ps} was carried out at a constant temperature of \SI{1}{K} to eliminate residual thermal fluctuations. Finally, a new energy minimization stage was conducted using the same parameters as in the initial stage to ensure geometric stability before DFT optimization.

Throughout the simulation process, Newton's equations of motion were integrated using the Velocity-Verlet algorithm \cite{swope1982computer} with a time step of \SI{0.1}{fs}. Interactions between carbon and hydrogen atoms were described using the Adaptive Intermolecular Reactive Empirical Bond Order (AIREBO) potential \cite{stuart2000reactive}. Temperature control was implemented through the Nosé-Hoover thermostat \cite{nose1984unified,hoover1985canonical}.

To simulate the interaction between OQCs and a metallic substrate, as occurs in the experimental synthesis, the same systems previously relaxed in the gas phase were repositioned at a distance of \text{3} \AA \ from the edge of a modified simulation box. This new configuration included non-periodic boundaries and a static substrate modeled through a 12-6 Lennard-Jones (LJ) potential \cite{yang2024unveiling}, given by
\begin{equation}
E_\text{C-Au} = 4\varepsilon_\text{C-Au} \left[ \left( \frac{\sigma_\text{C-Au}}{r} \right)^{12} - \left( \frac{\sigma_\text{C-Au}}{r} \right)^{6} \right],
\end{equation}
where $r$ is the distance between each atom and the substrate, $\varepsilon_\text{C-Au}$ is the potential well depth, and $\sigma_\text{C-Au}$ is the equilibrium distance. The interaction is truncated at a cutoff radius $r_C$, such that $E_\text{C-Au} = 0$ for $r > r_C$. For the interaction between carbon and gold, we used $\varepsilon_\text{C-Au} = \text{31.81}$ meV and $\sigma_\text{C-Au} =  \text{2.99}$  \AA  \ \cite{brann2021differential}. The interaction between hydrogen and gold was assumed to be negligible. After the classical simulations, the optimized OQC geometries were subjected to a new structural refinement via DFT, as described in the next section.

\subsection{Density Functional Theory} The geometric optimization and electronic structure of the OQCs were determined using first-principles calculations based on DFT, implemented in the SIESTA code. This code was selected for its efficiency in handling large-scale systems. The structural optimization was conducted within the generalized gradient approximation (GGA), using the Perdew-Burke-Ernzerhof (PBE) functional, widely recognized for its balance between computational cost and accuracy in describing atomic structures.

To mitigate the well-known underestimation of the band gap associated with GGA functionals \cite{xiao2011accurate}, we performed additional calculations with the hybrid functional HSE06 \cite{heyd2003hybrid,Heyd2006}, implemented through the HONPAS package \cite{Qin2015,Shang2020}. Brillouin zone integration was performed with a Monkhorst-Pack grid \cite{monkhorst1976} $1\times1\times10$ \textbf{k} along the periodic direction of the chains ($x$). The electron-ion interactions were described by norm-conserved Troullier-Martins pseudopotentials \cite{troullier1991,hamann1979}, in the factorized Kleinman-Bylander form \cite{kleinman1982}.

We used a 400 Ry grid cutoff for the kinetic energy and a single-$\zeta$ (SZ) basis. Structural optimizations involved complete relaxation of atomic positions and lattice vectors, with convergence criteria of strength of 0.01 eV/\r{A} and energy of $10^{-4}$ eV. To avoid spurious interactions between periodic images, periodic boundary conditions were applied exclusively in the axial direction of the chains, while vacuum regions larger than \text{35} \AA and \text{50} \AA were established in the transverse ($x$) and normal ($y$) directions, respectively.

\subsection{Tight-binding} The interplay of conjugation, symmetry, and quantum confinement effects governs the electronic properties of OQC systems. The quasi-one-dimensional nature of OQC leads to the formation of single-quantum-channel transport along the chain. The OQCs can be described using the tight-binding Hamiltonian,
\begin{equation}
H = \sum_{i} \epsilon_i c_i^\dagger c_i + \sum_{i,j} t_{ij} \left( c_i^\dagger c_j + \text{H.c.} \right),
\end{equation}
where $\epsilon_i$ represents the on-site energy, $t_{ij}$ is the hopping integral between sites $i$ and $j$, and $c_i^\dagger$ ($c_i$) is the creation (annihilation) operator for electrons at site $i$.

Quantum transport properties of OQCs are investigated using the Green function formalism. The Green function $\mathcal{G}(E)$ is defined as
\begin{equation}
\mathcal{G}(E) = \left[ E I - H - \Sigma_L(E) - \Sigma_R(E) \right]^{-1},
\end{equation}
where $E$ is the energy and $\Sigma_L(E)$ and $\Sigma_R(E)$ are the self-energies of the left and right electrodes, respectively. The total density of states (DOS) is obtained by $\text{DOS}=-(1/\pi)\text{Im} \{ Tr [ \mathcal{G}(E) ] \}$, \textcolor{black}{which is directly related to the summation of Bloch states over all crystal momenta $k$}. The transmission function $\mathcal{T}(E)$ is given by \cite{Datta},
\begin{equation}
\mathcal{T}(E) = \text{Tr} \left[ \Gamma_L(E) \mathcal{G}(E) \Gamma_R(E) \mathcal{G}^\dagger(E) \right],
\end{equation}
where $\Gamma_L(E)$ and $\Gamma_R(E)$ are the broadening matrices of the left and right electrodes, respectively. To pursue the spatial distribution of electronic transmission, we employ a local transmission coefficient $T^{\alpha}_{ij}(E)$ which computes the transmission between $j$ and $i$ sites, and leads $\alpha=L$ or $R$ \cite{PhysRevB.106.245408}, namely
\begin{equation}
    T^{\alpha}_{ij}(E)=2\text{Im}\{ [ \mathcal{G}(E)\Gamma_\alpha \mathcal{G}^\dagger(E) ]_{ji}H_{ij}\} \ .
\end{equation}
Local transmission provides the spatial information of the preferential paths for electrons in a given energy state. A high density in the local transmission implies in a higher probability of the state percolation across the atomic bonds.

\begin{figure*}[t!]
    \centering \includegraphics[width=18cm]{nFIG2.jpg}
\caption{
Electronic structure evolution of Organic Quantum Chains.
(a) Tight-binding band structures (red curves) for OQC configurations ranging from $N_1$ to $N_{10}$. For the $N_1$ case, DFT results obtained with the PBE (blue dotted lines) and HSE06 (solid black lines) functionals are also shown, together with the TB parametrization fitted to the HSE06 band structure. (b) DFT-PBE band structures for chains $N_1$ to $N_5$, confirming the systematic band splitting with increasing unit-cell size while preserving the central energy gap ($\Delta E_{\text{TB}} \approx 2.0$~eV and $\Delta E_{\text{PBE}} \approx 1.30$~eV), in qualitative agreement with the TB results. The TB fitting parameters are $t_1=-2.7$~eV, $\varepsilon=-0.55$~eV, and $1.85 \leq \beta \leq 2.0$. (c) Top: Zoom-in of the valence-band region for selected chains, highlighting the evolution of the valence-band maximum and the formation of sub-band groups, separated by dashed gray lines. Bottom: Projected LDOS associated with the valence-band maximum, mapped onto the atomic structures, revealing the spatial confinement and hierarchical localization of charge carriers within specific atomic clusters.
}
\label{fig:bandsldos}
\end{figure*}

\begin{figure*}[t!]
    \centering \includegraphics[width=16cm]{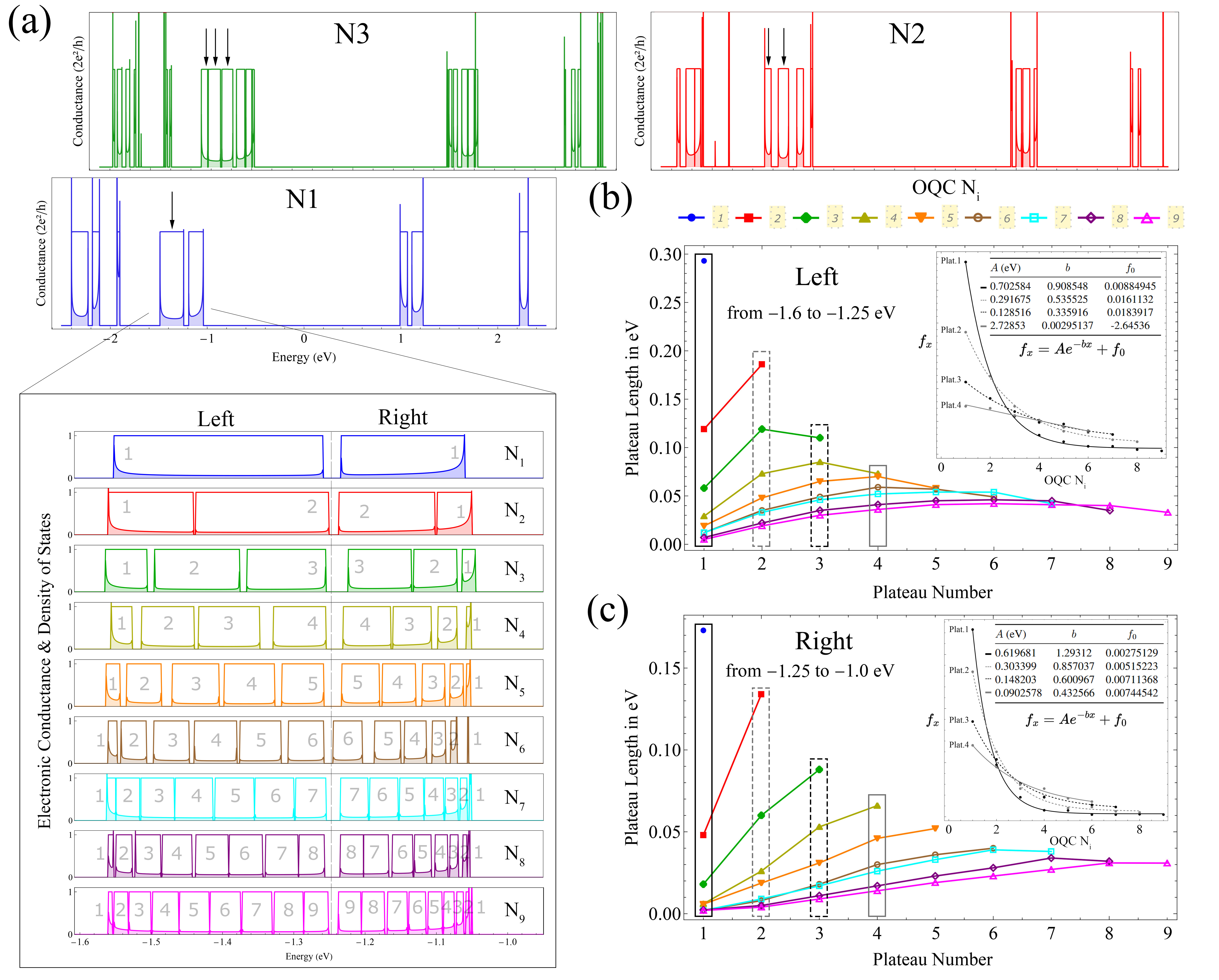}
    \caption{ Emergence of hierarchical conductance channels in OQCs. (a) Top: Conductance [$G (E)$, solid lines] and DOS (shaded areas) for chains $N_1$, $N_2$, and $N_3$. Black arrows highlight the splitting of a single molecular resonance ($N_1$) into $N$ distinct sub-levels due to inter-precursor hybridization. Bottom: Evolution of the conductance hierarchy for generations $N_1$ to $N_9$ within the representative mini-band at the valence states ($-1.6$ eV to $-1.0$ eV). Numbered blocks track the splitting of distinct conductance plateaus, divided into ``Left'' and ``Right'' spectral regions. (b, c) Evolution of the plateau length (energy width) as a function of the plateau index for the (b) Left and (c) Right spectral regions. The color  code corresponds to the $N_i$-OQC generation shown in (a). Insets: Exponential decay fits of the plateau length vs. generation index $N_i$ for the first four primary plateau families, as shown by the corresponding rectangles, quantifying the spectral densification.      }\label{fig:conductance_hierarchy}
\end{figure*}

The tight-binding parameters used in the whole calculations are obtained by fitting the electronic band structures with our DFT results. We employ the exponential decaying hopping parametrization, given by
\begin{equation}
    t_{ij}=t_1e^{-\beta\big(\frac{r_{ij}}{d_{\text{min}}}-1\big)},
\end{equation}
with $r_{ij}$ being the distance between $i,j$ lattice sites, $t_1$ the hopping related to the first nearest-neighbor distance $d_{\text{min}}= \text{1.37}$ \AA, and $\beta$ a fitting parameter that controls the range of the hopping energies. As the ratio $r_{ij}/d_{\text{min}}$ is always larger than one beyond the first nearest-neighbors, small values of $\beta$ increase the number of neighbors with non-negligible hopping contributions in the description. The nearest-neighbor distance between atoms inside the unit cell (black bounds in Fig.~\ref{fig:setup}) is within the range 1.37–1.50~\AA, obtained from the relaxed structure. This parametrization scheme has been successfully employed for carbon-based mixed geometries in graphene allotropes, such as multidimensional biphenylene systems \cite{lage2024,Lage2026}.

For the chains, the Green function formalism is employed, utilizing the decimation method to achieve numerical periodicity in momentum space. However, for finite systems, such as carbon molecule rings, the local density of states (LDOS) is calculated as a function of energy by
\begin{equation}
    LDOS(x,y,E)=\sum_n | \psi_n(x,y)|^2 \delta(E-E_n),
\end{equation}
where $x$ and $y$ are site coordinates and $\psi_n(x,y)$ is the corresponding electronic wave function of the $n^{th}$ state. \textcolor{black}{The real-space projection shown in our figures represents the total spectral weight accumulated at site $i$
for energy $E$.} To compute the delocalized electronic contribution across the system, given by $|\psi_n(x,y)|^2$, we transform the delta functions into Lorentzian functions, with $\Gamma$ fitted with the experimental results \cite{Nature2021}.

\section{Results}

\subsection{Organic Quantum Chains} The systems analyzed in this study exhibit distinct geometrical characteristics arising from the intrinsic flexibility of their carbon-hydrogen chains and the resulting curvature. The STM image shown in Fig.~\ref{fig:setup}(a) reveals a variety of configurations, including curved and closed structures, as well as linear segments that connect repeating units aligned along the surface. The internal cavities formed by the spatial arrangement of these units vary in size, leading to a porous and directionally anisotropic morphology.

Figures~\ref{fig:setup}(d-m) present the morphological evolution of the OQC structures as a function of the size of building blocks $N_i$, culminating in extended architectures. In particular, the configuration denoted as $N_{10}$ [see Fig.~\ref{fig:setup}(m)] exhibits lateral continuity along the $z$ direction, with a characteristic length $L_\infty$ that indicates a periodic vertical arrangement. 
Despite local variations in cavity size and curvature, the system maintains a well-defined periodicity along the axis of growth.

\textcolor{black}{The OQC periodicity along the z-direction ($L_\infty$) shown in Fig.~\ref{fig:setup}(d) illustrates the periodic chain for the $N_1$ case, where the gray area highlights one choice for the unit cell. The fundamental unit building blocks adopted to construct other OQCs are also illustrated in Fig.~\ref{fig:setup}(d) at the detached rectangle. The optimized structures obtained allow us to model not only the connectivity but also the structural evolution of the OQCs' curvature. For $N_2$-N$_9$, the translational unit cell contains the entire sinusoidal segment, capturing the strain compensation mechanisms that stabilize the complex organic nanostructures. The scattering setup adopted for the transport analysis of the OQCs is depicted in Fig.~\ref{fig:setup}(e) for the $N_{10}$-OQC case. The central region of electronic scattering is connected with the semi-infinite left and right leads.}

Building on the structural understanding and based on the energetic analysis detailed in the Supplementary Material (SM) \cite{SM}, we now examine how the electronic properties evolve along with the $N_{i}$ cases here explored. 

\textcolor{black}{The calculated band structures shown in Fig.~\ref{fig:bandsldos} reveal the intricate evolution of the electronic spectrum as the complexity of the unit cell increases from $N_1$ to $N_{10}$,} with length $a$ according to the OQCs. \textcolor{black}{In the panel corresponding to N$_1$-OQC, we present first-principles results obtained using both the DFT-PBE functional (blue dots) and the hybrid HSE06 functional (black lines). The TB parameters were chosen to accurately reproduce the upper valence and lower conduction bands obtained with the HSE06 potential. The entire sequence of TB band structures for larger unit cells (red curves) was then generated using the same parameter set.} 
The best TB parametrization adopted for all $N_i$-OQC structures in the present work are the following; $\beta=2.8/d_{\text{min}}=2.04$, $\epsilon =  -0.22$ eV, and $t_1= -3.4$ eV. The exponentially decaying hopping was limited by a cutoff radius of $R< {4.2}$ \AA.
\textcolor{black}{To validate the reliability of the TB model for extended systems with increasingly complex unit cells, we performed additional DFT-PBE calculations for configurations ranging from $N_1$- to $N_5$-OQC, as shown in Fig.~\ref{fig:bandsldos}(b). These results demonstrate that the TB parameterization consistently captures the essential physics as chain length increases, providing an efficient and accurate alternative when the computational cost of DFT calculations with complex functionals, such as HSE06, renders them practical. Both approaches reproduce the characteristic splitting of molecular levels and the persistence of the central energy gap ($\Delta E \approx 2.0$ eV, as obtained from the HSE06 fit).}

\textcolor{black}{As the number of precursors inside the unit cell grows, the electronic bands change accordingly by systematically subdividing into a set of mini-band spectra following the discrete molecular orbitals of the individual building blocks. A detailed inspection of the valence bands, shown in the upper panel of Fig.~\ref{fig:bandsldos} (c), reveals the transition of dispersive states to flat bands as the OQC changes from $N_3$- to $N_9$-OQC, when considering, for instance, the valence band maximum (VBM), highlighted in green.  To illustrate the physical features of these flat bands, we compute the real-space-projected LDOS of the VBM of the proposed chains. As displayed at the lower part of Fig.~\ref{fig:bandsldos} (c), the electronic probability densities are not always uniformly distributed along the chain. Depending on the $N_i$-OQC length, they may exhibit spatial confinement at specific segments of the precursor units. In the case of dispersive bands, the electronic state is delocalized over the entire unit cell, like $N_{3}$-, $N_{4}$-, and $N_{5}$-OQC cases, while flat bands lead to states localized at atomic clusters, as observed for $N_{6}$-, $N_7$-, and $N_9$-OQC, with no percolation within the entire unit cell.}

\textcolor{black}{To further understand the electronic evolution of the OQCs, we have analyzed the transport signatures of the valence states. Fig.~\ref{fig:conductance_hierarchy}(a) illustrates the formation of a hierarchy of conductance plateaus within a representative mini-band as previously discussed in Fig.~\ref{fig:bandsldos}(c) with energy range $-1.6$ to $-1.0$ eV. We notice that the outermost plateaus, labeled as 1 at left and right, respectively, have a decreasing length as the system grows, with a hierarchy induced by the subdivision of the bands being proportional to the number of inserted precursors pairs in the unit cell. The physical origin of this hierarchy is the hybridization of the discrete states molecular orbitals from the precursor units. As shown in the top panels of Fig.~\ref{fig:conductance_hierarchy}(a), a single transmission plateau observed in $N_1$ splits into exactly two sub-channels in the $N_2$, and three in the $N_3$ case. This pattern generalizes for any $N_i$ chain.  The lower panel of Fig.~\ref{fig:conductance_hierarchy}(a) tracks this evolution up to $N_9$.}

\textcolor{black}{The emergent hierarchy does not arise from a stochastic increase in energy levels, but from a well-defined, structured subdivision of the spectrum. The process follows a self-similar pattern across different energy ranges, as exhibited in Fig.~\ref{fig:conductance_hierarchy} (a) with a systematic partitioning quantified in panels (b) and (c) of Fig.~\ref{fig:conductance_hierarchy}. By analyzing the evolution of plateau length for the first 3 families, defined as the energy width of a single transmission window, we observe a robust exponential decay with rate parameter $b$. The mathematical trend provides a quantitative parameter for the emergent hierarchy in such systems. Moreover, it reflects a regular pattern of spectral subdivision where the fixed energy bandwidth of the initial $N_1$-OQC resonance is iteratively subdivided into $N$ discrete sub-levels and can be predicted as the chain length grows.} These findings show that OQC systems, with their rich electronic behavior, hold significant promise for carbon-based nanodevice applications.

\subsection{Coupled Corrals to OQCs} 
\begin{figure}[!h]
\centering
\includegraphics[width=\linewidth]{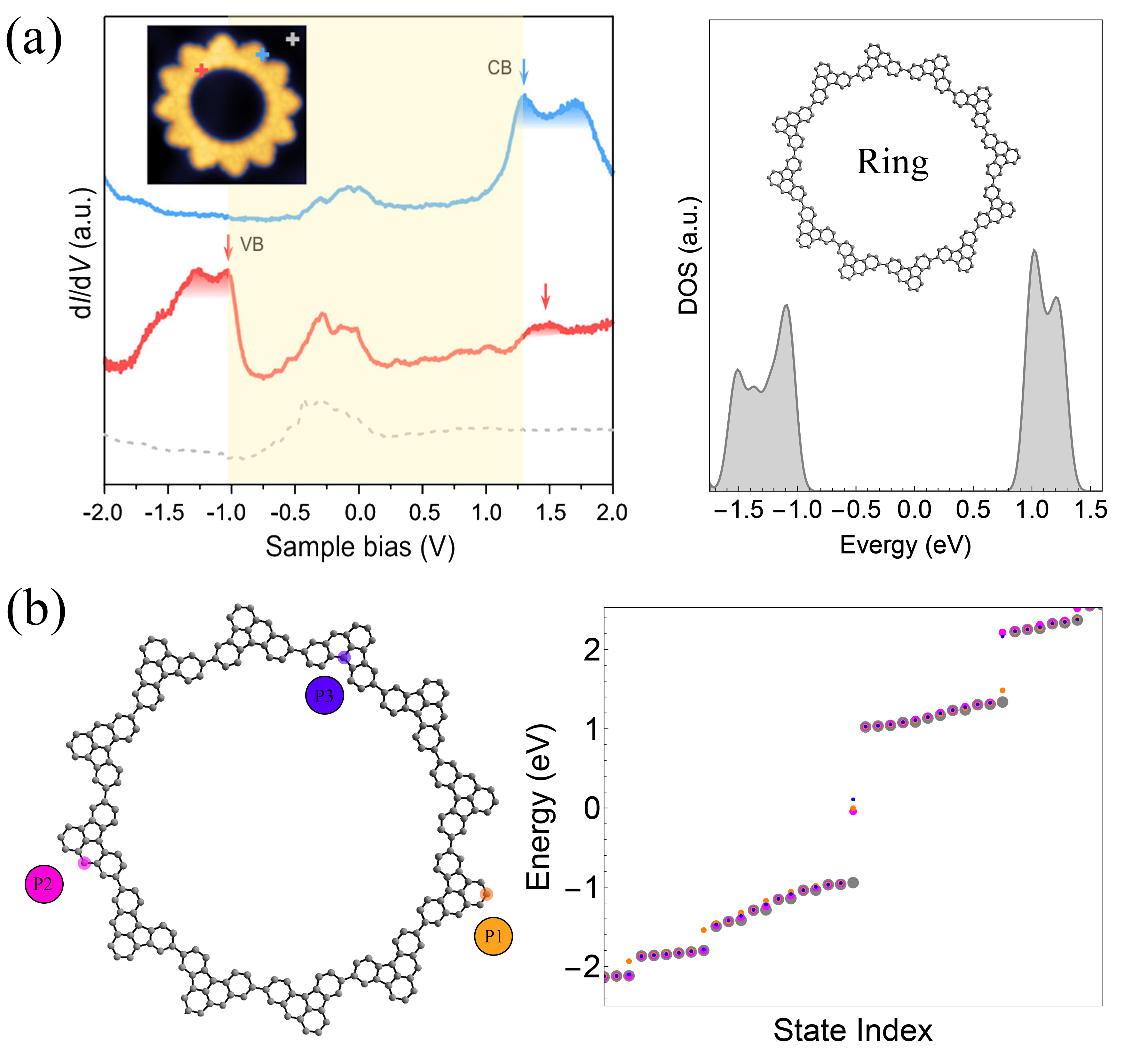}
    \caption{(a) Left: Experimental dI/dV spectrum of the quantum ring, exhibiting a fundamental gap of $\Delta E \approx 2.35$ eV (yellow shaded region) [Adapted~\cite{Nature2021}]. Right: Corresponding theoretical DOS for the pristine ring. (b) Left: Structural model showing the positions of coupled ions P1 (orange), P2 (magenta), and P3 (blue). Right: Discrete energy spectrum comparing the pristine ring states (gray circles) with the modified states induced by the ions. The colored markers explicitly show the emergence of localized in-gap states near the Fermi level and within the valence pseudo-gaps.}
   \label{fig:corral_dos}
\end{figure}

Apart from the isolated organic corral and chains exhibited in the experimental output shown in Fig.~\ref{fig:setup}(a), coupled quantum corrals and chains are also found in different configurations. Here, we investigate the transport and electronic responses of such systems. Most of the observed nanostructures  have $N_1$ chains coupled to a quantum ring by a single or a small set of precursors. 
We start by analyzing the total DOS of a single quantum corral as reported in the $dI/dV$ experimental results~\cite{Nature2021}, as shown in Figs.~\ref{fig:corral_dos}(a). The used tight-binding parametrization is the same as discussed before for the chains.
\textcolor{black}{ The experimental results show a fundamental gap of $\Delta E \approx 2.35$ eV (yellow region), while the TB model predicts a slightly smaller gap of $\Delta E \approx 2.0$ eV. This underestimation is typical for DFT-based parameterizations \cite{Borlido2020}. Notably, the experimental spectrum exhibits a broad mid-gap feature (dashed gray line) attributed to substrate hybridization. As our isolated model does not include the gold substrate, we investigate similar in-gap phenomena arising from local chemical hybridization by considering coupled ions at specific sites (P1, P2, P3) as shown in Fig.~\ref{fig:corral_dos}(b). The broken molecular symmetry introduces localized states deep within the fundamental gap (near $E=0$) and inside the valence pseudo-gaps. The discrete impurity levels (colored markers) effectively provide accessible quantum states which are often observed in substrate-coupled or doped molecular corrals. Features observed within the gap are attributed to Au substrate states rather than intrinsic molecular orbitals, being consistent with previous reports \cite{Nature2021,Feng2025,doi:10.1021/jacs.5c03736}.} We conclude that the sensitivity  of the rings to absorbed ions demonstrates the potential application within nanoscale sensors.

\begin{figure*}[t!]
    \centering
    \includegraphics[width=1.0\textwidth]{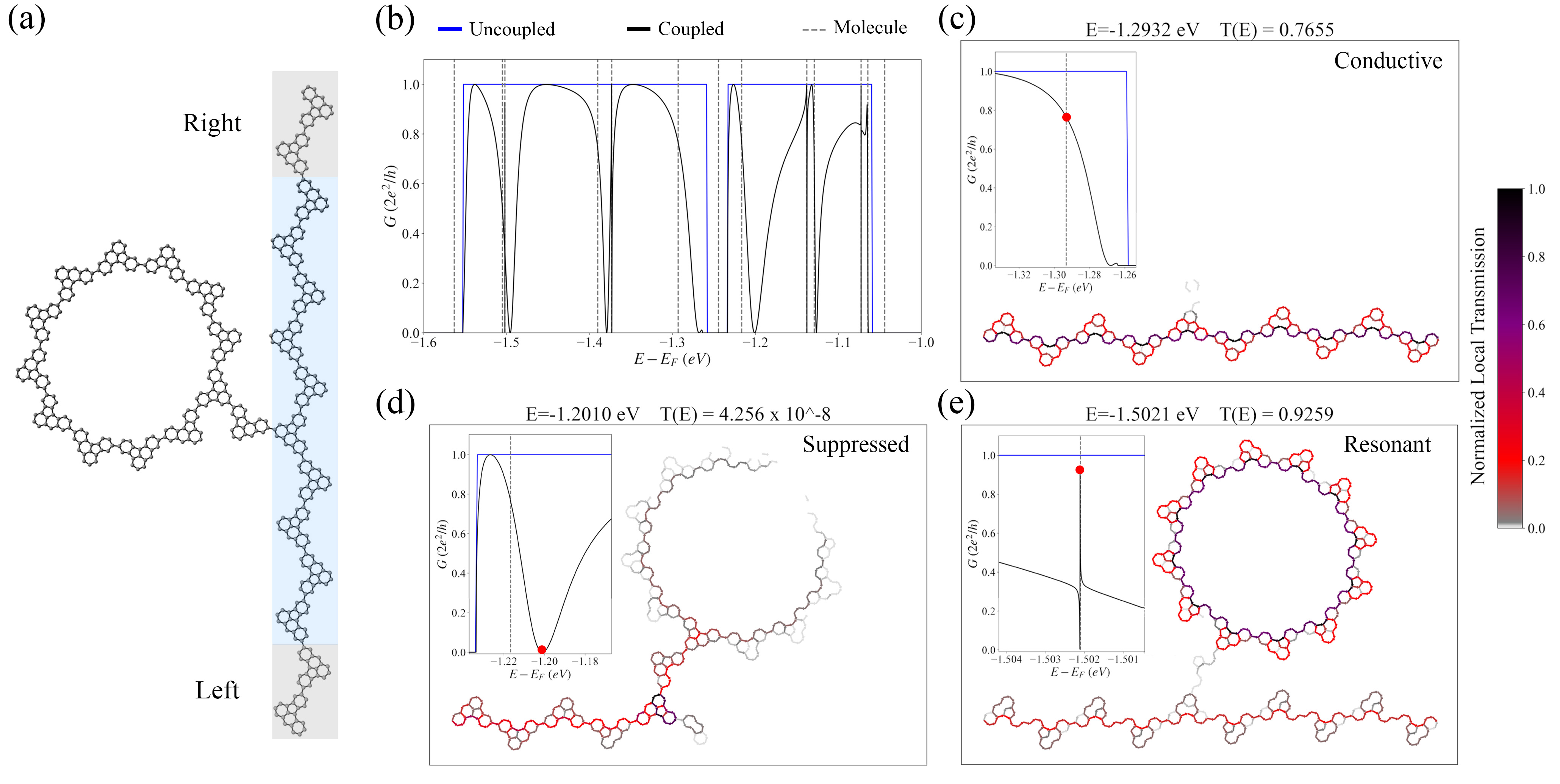}
    \caption{(a) Schematics of the scattering system. The gray shaded regions correspond to left and right leads connected to the scattering region shaded in blue. (b) Conductance for the uncoupled (blue curve) and coupled (black lines) systems. Vertical dashed gray lines mark the localized states of the molecule (the ring and one precursor). Local transmission for (c) Conductive, (d) \textcolor{black}{Suppressed}, and (e) Resonant cases, calculated at energies marked by red dots in the corresponding insets. Insets: conductance for the uncoupled and coupled systems and localized states of the molecule in dashed vertical lines.}
   \label{fig:transport_coupled}
\end{figure*}

The transport properties of the coupled system are explored by setting the scattering problem as shown in Fig.~\ref{fig:transport_coupled}(a), i.e. considering the left and right leads composed of $N_1$-OQC, and the center region formed by a piece of $N_1$-OQC attached to the molecule. Localized states of the molecule are represented by gray dashed vertical lines exhibited together with the conductance results for the uncoupled N$_1$-OQC and the coupled system composed of chain+precursor+corral, represented by blue and black curves, respectively, in Fig.~\ref{fig:transport_coupled}(b). The \textcolor{black}{molecule coupling} to the OQC affects the ballistic transport of the original $N_1$-OQC (blue curves) usually reducing the typical 1D channel of the $2e^2/h$ quantum \textcolor{black}{conductance}. However, at particular energy values, more dramatic changes are achieved, such as full conduction suppression or resonant state.

When spatially coupled, molecular states and the continuum states of the chain can coexist, giving rise to distinct transport responses, here classified as conductive, \textcolor{black}{suppressed}, and resonant, as illustrated by the normalized local transmission maps in Fig.~\ref{fig:transport_coupled}(c-e) with the insets highlighting the corresponding physical regimes. The local transmission patterns in panels (c) and (e) correspond to cases in which the local transmission is calculated exactly at a discrete molecular energy level (red dots in the conductance plot). In the resonant regime, electrons are confined to very narrow energy windows, exhibiting almost ballistic transport. In contrast, in the conductive regime, the molecular states are effectively absorbed into the continuum of the OQC.
In the latter case, the local transmission displays a preferential central route along the chain, although all chain bonds are effectively visited, in contrast to the corral sites, which do not contribute to transmission. Interestingly, in the resonant case the local transmission extends over both the chain and the  molecule, owing to a small but finite percolation \textcolor{black}{of the electronic states} through the precursor connecting the chain and the corral (gray shades). In fact, this contribution is further enhanced inside the coupled corral region, as indicated by the purple color scale in Fig.~\ref{fig:transport_coupled}(e).

\textcolor{black}{In general, several molecular localized states become embedded in the continuous of the coupled system (molecule+chain).} In this coupled regime, emergent resonant states of the finite system ($E_{res}$) \textcolor{black}{can give rise to strongly localized hybrid modes within the continuum, resulting in the formation of }bound states in the continuum (BIC). 
The intensity of this effect in electronic devices, may be quantified by a dimensionless quality factor, $  \mathcal{Q}=\frac{|E_{res}|}{\Delta E_{res} }$, where $\Delta E_{\mathrm{res}}$ denotes the width of the Fano antiresonance and characterizes the coupling-induced spectral broadening. The factor $\mathcal{Q}$  provides a measure of the quality of the resonant state,~\textcolor{black}{quantifying the degree of hybridization and the extent to which the sharp, discrete energy levels of the isolated molecule are absorbed into the continuum spectrum of the chain upon coupling}.  By definition, true BICs~\cite{orellana} correspond  to  the limit $\Delta E_{res} \rightarrow 0$, and consequently $\mathcal{Q}\rightarrow \infty$. For the resonant example shown in [Fig.~\ref{fig:transport_coupled}(e)]   the calculated quality factor indicates that the coupled state (inset black curve) nearly satisfies the condition $\Delta E_{res}\rightarrow0$, being defined as a quasi-BIC state. The molecular state (gray dashed line) gives rise to the resonance peak in the conductance, with $G(E)$ approaching to one quantum conductance flux, in close analogy with the continuum states of the chain (blue lines). 

For some energies of the injected current, zero \textcolor{black}{conductance is achieved} by the electronic states, due to destructive interferences that may occur inside the scattering region. As shown in the \textcolor{black}{suppressed} case in Fig.~\ref{fig:transport_coupled}(d), the state does not fully percolate through the central part, being spatially trapped between the left lead and the molecule.

\section*{Conclusions}

In this work, we have presented a theoretical analysis of experimentally synthesized organic quantum chains, combining DFT and CMD to obtain relaxed structures. The electronic properties were accurately captured by a TB model with an exponential hopping function within the ab initio relaxed structures, being in good agreement with DFT calculations.

A key consequence of the quasi-one-dimensional character of these systems is the formation of singular hierarchical conductance channels within the electronic spectra. The confinement is enhanced by inserting multiple precursors within the unit cell, leading to a partitioning of continuous states \textcolor{black}{and a densification of narrow conductance plateaus,} leading to a pseudo-band hierarchy.
\textcolor{black}{Our analysis revealed an interesting localization phenomenon governed by the molecular nature of the OQC building blocks. We identified confined molecular states near the Fermi level associated with electronic carriers trapped within specific precursor sub-units due to weak inter-unit coupling. These states appear as flat bands in the electronic spectrum and do not contribute to ballistic transport. }
This electronic structure makes these materials highly promising for multi-band light-emitting devices, as it promotes controlled electronic transitions between the emergent sub-bands. Furthermore, the molecular chains, are characterized by an scale invariant large energy gap of $\Delta E \approx 2.0$ eV for all proposed $N_i$ OQCs. This prediction matches very well with the measured experimental value \cite{Nature2021}.

Finally, we propose a quantum device based on the experimental formations of the carbon structures. The system composed of a OQC attached to a quantum corral shows prominent interferences and bound states in the continuum in the conductance results. For conductive situations, the \textcolor{black}{maximum} transmission probability occurs in the very center of the OQC, being completely suppressed along the coupled corral. Resonant cases appear as quasi-BICs with transmission occurring along both chain and molecule, being amplified inside the corral. 

We conclude that such systems are promising building blocks for the construction of carbon-based nano devices, exhibiting a combination of robust structural stability with highly tunable electronic properties. In particular, it has been possible to engineer a hierarchy of pseudo-gaps and induce localized states through smart molecular devices design.

\section*{Author contributions}
All authors contributed to the design of the research, the analysis of the results, and the writing of the manuscript. L.L.L. and A.B.F. performed the numerical calculations.

\section*{Conflicts of interest}
There are no conflicts to declare.

\section*{Data availability}

The code and files used within Tight-binding calculations can be found at \url{https://github.com/lucaslopeslage/OQC_code_and_files.git}.


\section*{Acknowledgements}

The authors  thank the INCT de Nanomateriais de Carbono for providing support on the computational infrastructure. LLL thanks the National Council for Scientific and Technological Development (CNPq)  scholarship. AL thanks FAPERJ under grant E-26/200.569/2023. D.S.G. acknowledges financial support from the Coordination for the Improvement of Higher Education Personnel (CAPES) under grant No. 88887.102348/2025-00 (Finance Code 001). M.L.P.J. acknowledges financial support from the Federal District Research Support Foundation (FAPDF) under grant No. 00193-00001807/2023-16 and from the CNPq under grant No. 444921/2024-9. The authors thank P. Orellana for fruitful discussions.

\bibliography{refs}

\end{document}